\documentclass[twocolumn,showpacs,preprintnumbers,amsmath,amssymb, superscriptaddress]{revtex4}

\usepackage{graphicx}
\usepackage{dcolumn}
\usepackage{bm}

\begin{document}

\preprint{APS/123-QED}

\title{Biphoton compression in standard optical fiber: exact numerical calculation}

\author{G.~Brida}
\affiliation{Istituto Nazionale di Ricerca Metrologica, Strada delle
Cacce 91, 10135 Torino, Italy}
\author{M.~V.~Chekhova}
\affiliation{Department of Physics, M.V.Lomonosov Moscow State
University,\\  Leninskie Gory, 119991 Moscow, Russia}
\affiliation{Istituto Nazionale di Ricerca Metrologica, Strada delle
Cacce 91, 10135 Torino, Italy}
\author{I.~P.~Degiovanni}
\affiliation{Istituto Nazionale di Ricerca Metrologica, Strada delle
Cacce 91, 10135 Torino, Italy}
\author{M.~Genovese}
\affiliation{Istituto Nazionale di Ricerca Metrologica, Strada delle
Cacce 91, 10135 Torino, Italy}
\author{G.~Kh.~Kitaeva}
\affiliation{Department of Physics, M.V.Lomonosov Moscow State
University,\\  Leninskie Gory, 119991 Moscow, Russia}
\author{A.~Meda}
\affiliation{Istituto Nazionale di Ricerca Metrologica, Strada delle
Cacce 91, 10135 Torino, Italy}
\author{O.~A.~Shumilkina}
\affiliation{Department of Physics, M.V.Lomonosov Moscow State
University,\\  Leninskie Gory, 119991 Moscow, Russia}

\date{\today}

\begin{abstract}
Generation of two-photon wavepackets, produced by spontaneous
parametric down conversion in crystals with linearly chirped
quasi-phase matching grating, is analyzed. Although being spectrally
broad, two-photon wavepackets produced this way are not Fourier
transform limited. In the paper we discuss the temporal compression
of the wavepackets, exploiting the insertion of a standard optical
fiber in the path of one of the two photons. The effect is analyzed
by means of full numerical calculation and the exact dispersion
dependencies in both the crystal and the fiber are considered. The
study opens the way to the practical realization of this idea.
\end{abstract}

\pacs{Valid PACS appear here}
\maketitle

\section{\label{sec:level1}Introduction}

One of the problems in modern quantum optics is generation of
two-photon light with given spectral and temporal properties. In
particular, of considerable interest is preparation of two-photon
wavepackets with small correlation times. Studies in this direction
are important, first of all, from the viewpoint of the fundamental
question: how well can we localize a photon of a correlated pair if
its twin is registered at a certain time?  Furthermore, two-photon
wavepackets with small correlation times are required for quantum
technology applications involving interactions between matter and
nonclassical light \cite{1m, 1m2, H41,H42, H5}, as well as for
quantum metrology applications \cite{2m}. Several ideas have been
suggested for generating two-photon states with broad spectra, all
based on Spontaneous Parametric Down Conversion (SPDC). Among them,
one can mention prisms or diffraction gratings introducing a
frequency chirp \cite{1}, SPDC in aperiodically poled crystals
(APPC) \cite{2,Harris,4} or in crystals with temperature gradients
\cite{5}. However, a broad spectrum of two-photon light does not
necessarily lead to small correlation times, although the inverse is
true \cite{6,7,8}. Similarly, in classical optics, a broadband pulse
is not always short in time, although the spectrum of a short pulse
is always broad. Indeed, the spectrum broadening introduced in Refs.
\cite{1,4,5} is inhomogeneous, in the sense that two-photon
wavepackets generated in different parts of the crystal have
different spectra. As a result, the two-photon spectral amplitude
has a phase (a frequency chirp) depending nonlinearly on the
frequency and making the two-photon wavepackets not Fourier
transform-limited. Therefore, in these examples the biphotons are
not short in time despite their broad frequency spectrum. In Ref.
\cite{Harris} it was mentioned that time compression of such
two-photon wavepackets requires compensation for their frequency
chirp, although the way to eliminate the chirp was not specified.
Recently we have demonstrated  \cite{Nostro} that broadband biphoton
wave packets produced via SPDC in APPC can be compressed in time by
using group-velocity dispersion (GVD) in standard optical fiber. A
Fourier limited compression was obtained in \cite{Nostro} via an
analytical calculation within the approximation of large difference
in the group velocities of the two SPDC photons and neglecting
third-order GVD of the fiber. Furthermore, we have given an example
of a complete numerical calculation demonstrating that higher orders
do not change this result.

 With the purpose of paving
the way for a realistic experimental implementation of this idea and
application to quantum technologies \cite{mg}, here we present and
discuss a systematic study of this phenomenon. Furthermore, we
present some optimal situation where the full numerical calculation
shows an effect that can be clearly observed with a realistic
set-up.

\section{\label{sec:level2}Spectral properties of biphotons
emitted in a medium with second-order susceptibility varying along
\textbf{the pump propagation}}

Spontaneous Parametric Down Conversion occurs in non-linear
dielectric crystals and consists of the spontaneous decay of a
photon (pump) of frequency $\omega_p$ in two photons of lower
frequencies $\omega_s$ and $\omega_i$, historically named signal and
idler, satisfying, in an ideal case, the constraints of energy and
momentum conservation. In what follows we will consider type II
collinear SPDC
 interaction, with the pump being a linearly (ordinarily) polarized monochromatic
 plane wave propagating along the z direction and
the signal and idler fields being extraordinarily and ordinarily
polarized, respectively. Considering signal and idler fields
initially in the vacuum state, the output state is
\begin{equation}\label{state2}
|\psi\rangle = |0\rangle+\int d\omega_s d\omega_i
F(\omega_s,\omega_i)\hat{a}^{\dag}_{\omega_s}\hat{b}^{\dag}_{\omega_i}|0\rangle,
\end{equation}
where, due to the weak interaction, only the first two terms of the
perturbative expansion are considered \cite{Mandel}. In Eq.
(\ref{state2}), $\hat{a}^{\dag}_{\omega_s}$ and
$\hat{b}^{\dag}_{\omega_i}$
 are the photon creation operators for signal and idler photons, respectively.

 The function $F(\omega_s,\omega_i)$ is the Two-Photon
Spectral Amplitude (TPSA) and accounts, in particular, for the
non-linearity of the crystal $\chi^{(2)}$ and the wavevector
mismatch $k_p-k_s-k_i$. The second-order susceptibility is assumed
to depend on the pump propagation direction z only; for cw pumping,
the TPSA is

\begin{equation}\label{TPA}
F(\omega_s,\omega_i)\propto\delta(\omega_{p}-\omega_s-\omega_i)\int_{-L}^{0}dz
\chi^{(2)}(z)e^{i(k_p-k_s-k_i)z}, \\
\end{equation}
where \textit{L} is the length of the crystal and the wavevectors
$k_j=\frac{n_j(\omega_j)\omega_j}{c}$ ($j=p,s,i$) depend on the
ordinary or extraordinary refractive indexes $n_j$ of the crystal
\cite{Klyshko e Weinberg}.
 The TPSA determines all the
spectral and temporal properties of two-photon light. In the general
form, its square module gives the spectrum of the biphoton
radiation. For example, the signal spectrum is
\begin{equation}\label{Spect} S(\omega_s)=\left| \int d\omega_i
F(\omega_s,\omega_i)\right|^2,
\end{equation}
and similarly for the idler spectrum.


In bulk crystals the momentum conservation is ensured, for each
frequency, by choosing a proper direction of the waves involved in
the process. Nevertheless, in type II SPDC the efficiency of the
process is reduced by birefringence walk-off. The advantage of using
 poled crystals for biphoton generation relies on the absence
of transverse walk-off and allows one to avoid using narrowband
filtering in certain experiments. Moreover, compared with a bulk
crystal, a poled material provides a higher SPDC pair production
efficiency due to noncritical quasi-phase matching (QPM), obtainable
at a given temperature, or due to the possibility to utilize the
largest value of the effective nonlinear coefficient.

According to \cite{fejer}, let us consider a square nonuniform
first-order QPM grating with slow variation of the poling period
$\Lambda(z)$.
Then the spatial variation of the second-order
susceptibility along the medium can be written as
\begin{equation}\label{chi}
\chi^{(2)}(z)=\chi^{(2)}_0 e^{i K_0 z+i \Phi_0(z)},
\end{equation}
where
the absolute value of the second-order
susceptibility is constant.
The phase $\Phi_0(z)$ in Eq. (\ref{chi}) is induced by the
modulation of the local wavevector $\kappa(z)= 2 \pi / \Lambda(z)$.
The wavevector varies linearly within the length of the crystal
(linearly chirped QPM grating), thus \cite{fejer}
\begin{equation}\label{phi_chi}
\kappa(z)=K_0+\frac{d\Phi_0(z)}{dz}=K_0-2 \alpha(z+\frac{L}{2}),
\end{equation}
where $\alpha$ is defined as the chirping parameter and $K_0$ is
chosen so that perfect collinear first-order quasi phase matching
for selected frequencies $\omega_{s0}$ and $\omega_{i0}=
\omega_{p}-\omega_{s0}$ is obtained at the center of the crystal,
i.e.,
\begin{equation}\label{pm0}
k_p(\omega_{p})-k_s(\omega_{s0})-k_i(\omega_{p}-\omega_{s0})-K_0 =
0.
\end{equation}
The phase of the grating, obtained by integrating Eq.
(\ref{phi_chi}) (except for an unessential constant phase factor),
is
\begin{equation}
\Phi_0(z)=-\alpha(z+\frac{L}{2})^2
\end{equation}
and it equals zero at the center of the crystal.

The TPSA gains a phase shift that depends nonlinearly on the
frequency because of the presence of this phase factor in the
second-order nonlinear coefficient. In fact, doing analytically the
integral in Eq. (\ref{TPA}), we find that the TPSA takes the form

\begin{eqnarray} \label{TPAErf}
F(\omega_s,\omega_i) \propto e^{i \phi(\omega_s,\omega_i)}\nonumber\\
\times\left(\frac{\mathrm{Erf}[\frac{(-1)^{1/4}(L \alpha -\Delta
k)}{2 \sqrt{\alpha}}]}{\sqrt{\alpha}}+
\frac{\mathrm{Erf}[\frac{(-1)^{1/4}(L \alpha +\Delta k)}{2
\sqrt{\alpha}}]}{\sqrt{\alpha}}\right)\nonumber\\
\times\delta(\omega_{p}-\omega_s-\omega_i)
\end{eqnarray}
where $\mathrm{Erf}$ is the error function and $\Delta
k(\omega_s,\omega_i) = k_p-k_s-k_i-K_0$. The phase factor shows a
nonlinear dependence on $\Delta k(\omega_s,\omega_i)$:

\begin{equation} \label{phasefact}
\phi(\omega_s,\omega_i)=\frac{\Delta
k(\omega_s,\omega_i)L}{2}+\frac{\Delta k^2(\omega_s,\omega_i)}{4
\alpha}.
\end{equation}
As a consequence, the phase depends nonlinearly on the frequency, as
discussed in  more detail in section \ref{sec:level2}. This phase
factor does not affect the spectrum of the biphoton wavepackets as
it is evident from (\ref{Spect}). Nevertheless, it is of huge
importance when considering the temporal behavior of the wavepacket
\cite{Harris}. On the contrary, the error function in Eq.
(\ref{TPAErf}) affects the width of the spectrum of biphotons and
the effect is the broadening of the spectrum when increasing the
aperiodicity.

\section{\label{sec:level2}Temporal properties}
\begin{figure} \includegraphics[width=0.5\textwidth]{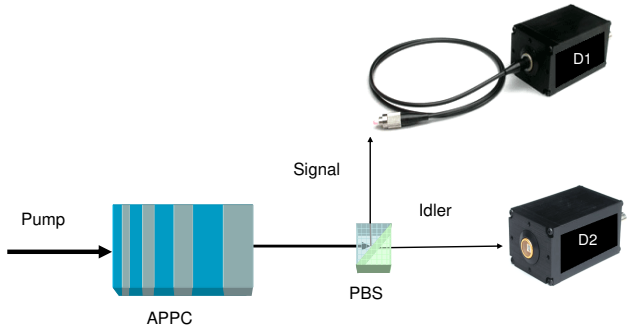}
\caption{(color online) A typical set-up for observing two-photon
coincidences in type-II SPDC. A laser beam pumps a type-II
aperiodically poled crystal (APPC) in collinear regime. After a
polarizing beam splitter (PBS), signal and idler beams are addressed
to low-jitter photodetectors, whose outputs are sent to a
coincidence circuit. \label{Coinc}}
\end{figure}

In Fig. \ref{Coinc} a typical experimental set-up for the
measurement of the time separation between the arrivals of the two
photons, $\tau$, is depicted. Photons of a pair produced by type II
SPDC are orthogonally polarized, and can be easily separated by a
polarizing beam-splitter and addressed to two different photon
counting detectors, D1 and D2.

The rate of coincidences between signal and idler photons reaching
detectors at times $t^{'}$ and $t^{''}$, respectively, can be
expressed as \cite{7}
\begin{equation}\label{RC}
R_c\propto\iint dt^{'} dt^{''} G_{meas}^{(2)}(t^{'};t^{''})
\Pi(t^{'}-t^{''}, T_W),
\end{equation}
where $\Pi(t^{''}-t^{'}, T_W)$ is the rectangular coincidence window
function of temporal width $T_W$ and $G_{meas}^{(2)}(t^{'};t^{''})$
is the second-order Glauber's correlation function, where, for
convenience, not only the electric field operators are incorporated,
but also the spectral response functions of the signal and idler
detection arms, $\mathfrak{f}_s$ and $\mathfrak{f}_i$, respectively.
$G_{meas}^{(2)}(t^{'};t^{''})$ stands then for the measured value of
Glauber's correlation function. Then,
\begin{eqnarray}\label{G2}
G_{meas}^{(2)}(t^{'};t^{''})&=&\langle\psi|\hat{A}_s^{\dag}(t^{'})\hat{B}_i^{\dag}(t^{''})\hat{B}_i(t^{''})\hat{A}_s(t^{'})|\psi\rangle
\\
\nonumber&=&\| \langle 0
|\hat{B}_i(t^{''})\hat{A}_s(t^{'})|\psi\rangle \|^2,
\end{eqnarray}

where

\begin{eqnarray}
\nonumber
 \hat{A}_{s}(t^{'})=\int d\omega_s
\mathfrak{f}_s(\omega_s)e^{-i\omega_s t^{'}} \hat{a}_{\omega_s},
\\
\nonumber \hat{B}_{i}(t'')=\int d\omega_i
\mathfrak{f}_i(\omega_i)e^{-i\omega_i t^{''}} \hat{b}_{\omega_i}.
\end{eqnarray}
Typically in the literature (see for example Ref. \cite{Mandel})
there is no spectral dependence of the detection arms, and the
operators $A_s(t^{'})$ and $B_s(t^{''})$ are usual quantized
electric fields, i.e.,
\begin{equation}\label{spectr_flat}
\mathfrak{f}_x(\omega_x) \varpropto 1,
\end{equation}
with $x=s,i$. However, in the general case the spectral response of
the detection channel should be taken into account since it
influences both spectral and temporal properties of the measured
two-photon wavepackets. For instance, considering Gaussian spectral
responses of the detection channels, we have
\begin{equation}\label{gauss_fil}
\mathfrak{f}_x(\omega_x) \varpropto
e^{-\frac{(\omega_x-\omega_{xc})^2}{(\delta \omega)^2}},
\end{equation}
where $\omega_{xc}$ is the selected central frequency of the
spectral response and $\delta \omega$ is its width. According to
Ref. \cite{rubin}, the measured time two-photon amplitude (TTPA) is
expressed as
\begin{eqnarray}
\label{state_after} \langle 0|B_i(t^{''}) A_s(t^{'})|\psi\rangle=
\int d\omega_s d\omega_i
F(\omega_s,\omega_i)\mathfrak{f}_s(\omega_s)\mathfrak{f}_i(\omega_i)\nonumber\\
\times e^{-i \omega_s t^{'}}e^{-i \omega_i t^{''}}.
\end{eqnarray}
The measured spectrum (\ref{Spect}) of the wavepacket is also
modified by the detection arms spectral filtering as
\begin{equation}\label{Spect_det}
S(\omega_s)=\left|\int d\omega_i
F(\omega_s,\omega_i)\mathfrak{f}_s(\omega_s)\mathfrak{f}_i(\omega_i)\right|^2.
\end{equation}
Until the end of the next section we will consider only the simplest
case of detection arms without any spectral dependence, i.e.,
$\mathfrak{f}_x(\omega_x)=1$, in order to focus on the temporal
properties of the emission. The discussion on the modification of
the temporal properties induced by the spectral characteristics of
the detection arms is postponed to section \ref{sec:level5}. To
discuss the difference in the arrival times of the signal photon,
$t^{'}$, and the idler one, $t^{''}$, on the detectors, we introduce
the new variables:
\begin{eqnarray}\label{tau}
\tau &=& t^{'}-t^{''}
\\
\nonumber T &=& t^{'}+t^{''}.
\end{eqnarray}
Thus, the TTPA can be rewritten as the Fourier transform of the
TPSA:
\begin{eqnarray}\label{state_after2}
\nonumber F(\tau)&=&\langle 0|B_i(t^{''}) A_s(t^{'}) |\psi\rangle
\\
\nonumber &\varpropto& \int d \omega_s F(\omega_s,
\omega_{p}-\omega_s) e^{i \omega_s \tau}.
\end{eqnarray}

The square module of the TTPA is the measured Glauber's second-order
correlation function,
\begin{equation} \label{g2ttpa}
G_{meas}^{(2)}(\tau)=|F(\tau)|^2
\end{equation}
and the width of $G_{meas}^{(2)}(\tau)$ will be further referred to
as correlation time.

\section{\label{sec:level4}Temporal biphoton compression}

 Since the TTPA is the Fourier transform of TPSA, a
necessary condition for reaching ultra narrow correlation time is
producing ultra broadband SPDC light. In the case of aperiodically
poled crystals with linear chirping, the phase factor depending
nonlinearly on $z$ is responsible for the spectral broadening effect
and increasing the aperiodicity broadens the spectrum.
Unfortunately, the two-photon spectral amplitude is not Fourier
transform limited because of the presence of the phase factor
(\ref{phasefact}) in Eq. (\ref{TPAErf}).

Thus, the spread in the detection time difference of the two photons
cannot be reduced simply by broadening the spectrum. As suggested in
Ref. \cite{Harris}, the ideal way to make TTPA perfectly Fourier
transform limited is to insert in the path of the biphotons a proper
optical medium that compensates for the nonlinear part of the phase
factor in Eq. (\ref{phasefact}). In general the insertion of an
optical medium with refractive index $n_{mx}(\omega)$ and of length
$l_x$ in both optical paths $x= s, i$ adds to the TPSA the phase
factor $e^{i (k_{ms}(\omega_s)l_s+k_{mi}(\omega_i)l_i)}$, where
$k_{mx}=\frac{n_{mx}(\omega_x)\omega_x}{c}$. Denoting the TPSA after
the propagation of biphotons through the optical medium as $F_m$, we
have

\begin{equation} \label{medium}
F_m(\omega_s,\omega_i)= e^{i (k_{ms}(\omega_s) l_s+k_{mi}(\omega_i)
l_i)}F(\omega_s,\omega_i).
\end{equation}

 In Ref. \cite{Harris} it was shown that a correlation time at the Fourier
 transform limit can be achieved when the inserted optical media satisfies the condition
\begin{equation} \label{pippo}
k_{ms}(\omega_s) l_s+k_{mi}(\omega_p-\omega_s) l_i = -\phi(\omega_s,
\omega_p-\omega_s).
\end{equation}

In Ref.~\cite{Nostro} it was shown that, in some specific cases,
this can be achieved by passing one of the two photon through a
normal-dispersion medium, for example a standard optical fiber. This
result opens a perspective of a real implementation of the method
suggested in Ref. \cite{Harris}. In particular, we considered the
simplest case where in Eq. (\ref{pippo}) the terms of orders higher
than two in the frequency expansion of the wavevector are
negligible. Here we explicitly discuss the effect of considering
such higher-order terms in the frequency expansion of Eq.
(\ref{pippo}). The first obvious consideration is that a perfect
Fourier transform limited compression does not require Eq.
(\ref{pippo}) to be satisfied; in fact, it is sufficient to satisfy
its second-order derivative:
\begin{equation} \label{der2}
\frac{d^2}{d \omega_s
^2}[k_{ms}(\omega_s)l_s+k_{mi}(\omega_p-\omega_s)l_i] = -
\frac{d^2}{d \omega_s^2}\phi(\omega_s,\omega_p-\omega_s).
\end{equation}
Furthermore, it is worth observing that Eq.(\ref{der2}) should be
satisfied only for the values of $\omega_s$ where the TPSA is
nonzero.


In order to understand how to satisfy, at least approximately, the
condition in Eq. (\ref{der2}) we compare the behavior of the
second-order derivative of the Harris Phase factor, $-\frac{d^2}{d
\omega_s^2}\phi(\omega_s,\omega_p-\omega_s)$ (HP), which needs to be
eliminated to provide the temporal compression of the biphoton at
the Fourier limit, and the one induced by the presence of the
optical fibers of appropriate lengths in the signal and idler
channels, $\frac{d^2}{d \omega_s
^2}[k_{ms}(\omega_s)l_s+k_{mi}(\omega_p-\omega_s)l_i]$ (OFP). We
considered collinear type II, almost degenerate, SPDC emission in an
aperiodically poled KTP crystal (APKTP). The monochromatic pump at
458 nm and the idler field are ordinarily polarized, while the
signal field is extraordinarily polarized. The dependence of
wavectors on the frequency in KTP is evaluated exploiting Sellmeier
equations ~\cite{sellmeier} and we consider the first-order QPM. The
phase mismatch at the center of the APKTP crystal is compensated for
by the inverse grating vector $K_0=2441.8$ cm$^{-1}$, corresponding
to degenerate emission at 916 nm. Spectral and temporal properties
of the SPDC radiation, as well as the temporal compression of the
biphoton wavepackets in the optical fiber, are numerically
calculated for several different parameters such as the length of
the crystal, the chirping parameter and the fiber length. No
frequency expansion of the wavevector is involved, in contrast to
Ref. \cite{Nostro} where the analysis was performed analytically and
only up to second-order terms in frequency.

\begin{figure}
\includegraphics[width=0.5\textwidth]{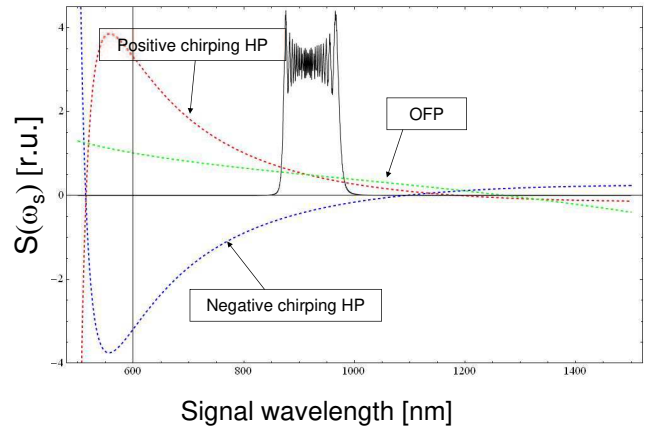}
\caption{(color online) The dependence of HP and OFP on the signal
wavelength (both for positive and negative chirping) with the
optical fiber present only in the signal channel, for a crystal with
$L = 0.8$ cm and $|\alpha| = 50 cm^{-2}$. For completeness, the
spectrum of the biphoton field is shown as well}. \label{GVDs}
\end{figure}

Fig. (\ref{GVDs}) shows the dependence of HP and OFP on the signal
wavelength in the case where the optical fiber is present only in
the signal channel, for a crystal with $L = 0.8$ cm and $|\alpha| =
50 cm^{-2}$ . For completeness, the spectrum of the biphoton field
is shown as well. It is interesting that the effect of the optical
fiber in either signal or idler SPDC channel is different for the
cases of positive ($\alpha
> 0$) and negative ($\alpha < 0$) chirping parameter.
In fact, since temporal compression is achieved due to
the group velocity dispersion in a standard optical fiber, only in
the case of negative chirping the HP can be approximately
compensated by the OFP. Temporal compression in the case of $\alpha
> 0$ can be obtained only in a specially engineered fiber with
negative group velocity dispersion, whose study is beyond the scope
of this paper.

\begin{figure}
\includegraphics[width=0.5\textwidth]{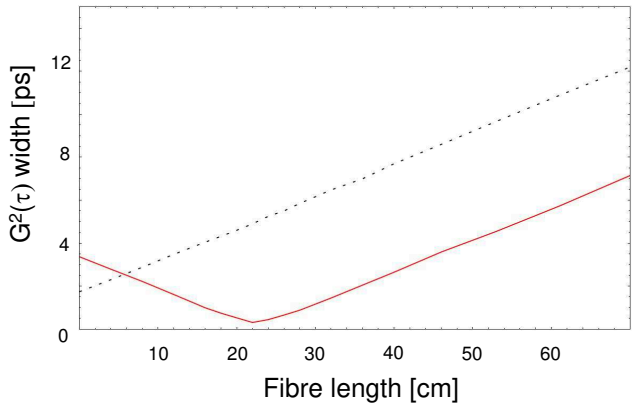}
\caption{(color online) Dependence of the correlation time on the
optical fiber length. The two lines refer to positive (dashed line)
and negative (solid line) chirping parameters. \label{Lines}}
\end{figure}

The maximum value of the compression is achieved for the length of
the fiber $l_f$ providing the best agreement between the HP and OFP,
approximately corresponding to the situation where HP and OFP have
the same value at the degenerate wavelength (916 nm), as shown in
Fig. \ref{GVDs}. The dependence of the correlation time on the
optical fiber length is shown in Fig. \ref{Lines} both for positive
 and negative  chirping parameter. The
effect of the optical fiber in the first case is an increase in the
correlation time, i.e., broadening of $G_{meas}^{(2)}(\tau)$, while
in the second case the insertion of the fiber leads to a compression
of $G_{meas}^{(2)}(\tau)$, until its minimum width is reached at
$l=l_f$. The compression is then followed by a broadening of
$G_{meas}^{(2)}(\tau)$.

At $l = 0$ the computed correlation times do not have the same
values for negative and positive chirp. This effect is due to the
generation of signal and idler photons at different frequencies at
the beginning and at the end of the crystal. In fact, biphotons are
degenerate only at the center of the crystal and the frequency
difference causes distinct delays between the two photons. During
the propagation of biphotons in the crystal, the delay due to the
frequency difference is compensated by the e-o delay in the case of
$\alpha>0$ but in the case of $\alpha<0$, the delays add up.

\begin{figure}
\includegraphics[width=0.5\textwidth]{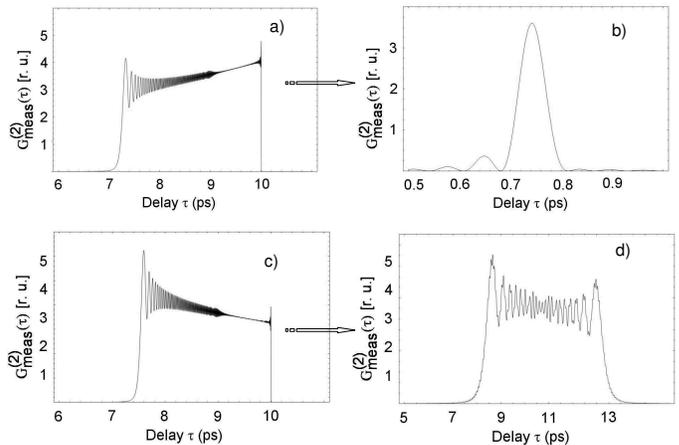}
\caption{(color online) $G_{meas}^{(2)}(\tau)$ evaluated via exact
numerical calculation for a crystal with $L = 0.8$ cm and $|\alpha|
= 20$ cm$^{-2}$. The left-hand part of the figure shows the shape of
$G_{meas}^{(2)}(\tau)$ at the output of the crystal, with no
dispersive medium inserted, while the right-hand part illustrates
the effect of a 1.06 m long fiber inserted into the SPDC signal
channel. The a) and b) parts refer to a crystal with $\alpha < 0$
while the c) and d) parts, to an $\alpha > 0$ crystal. \label{sinc}}
\end{figure}

\begin{figure}
\includegraphics[width=0.5\textwidth]{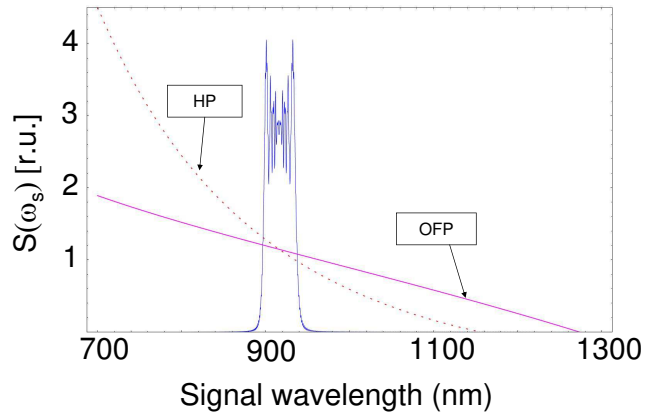}
\caption{(color online) The spectrum of biphotons, with the width of
approximately 40 nm, as well as HP and OFP frequency dependencies
(dashed and solid lines, respectively).\label{spect_sinc}}
\end{figure}

In \cite{Nostro} it was shown that within the second-order frequency
series expansion of $\phi(\omega_s,\omega_p-\omega_s)$, perfect
temporal compression of $G_{meas}^{(2)}(\tau)$ can be achieved with
normal dispersive media. Let us now discuss the effect of
higher-order terms in the expansion on the width of the compressed
$G_{meas}^{(2)}(\tau)$. Fig. \ref{sinc} reports the results obtained
via exact numerical calculation for a crystal with $L = 0.8$ cm and
$|\alpha| = 20$ cm$^{-2}$. The left-hand part of the figure shows
the shape of $G_{meas}^{(2)}(\tau)$ at the output of the crystal,
with no dispersive medium inserted, while the right-hand part
illustrates the effect of a 1.06 m long fiber inserted into the SPDC
signal channel. The a) and b) parts refer to a crystal with a
negative chirping parameter, $\alpha < 0$ , while the c) and d)
parts to an $\alpha > 0$ crystal. In the case of negative chirping,
the compressed $G_{meas}^{(2)}(\tau)$ takes almost the shape of an
ideal sinc function, meaning that time compression at the Fourier
limit is almost achieved \cite{Nostro}. The estimated FWHM of
$G_{meas}^{(2)}(\tau)$ is reduced from 2.8 ps to 67 fs (which is the
value at the Fourier limit), as it is clear from Fig.
\ref{spect_sinc}, where the spectrum of the corresponding biphoton
wavepacket is shown. Summarizing, Fig. \ref{sinc} shows that a
normal dispersive medium, as an optical fiber, can compress the
biphoton wavepacket to the Fourier limit. The effect is opposite in
the case of $\alpha > 0$: the initial correlation time of 2.6 ps is
increased to become 5 ps.

With this choice of crystal parameters, the condition given by
Eq.(\ref{der2}) should be satisfied in a quite narrow region of
nonzero signal field spectrum. Fig. \ref{spect_sinc} shows the
spectrum of  biphotons, approximately 40 nm wide, as well as HP and
OFP lines: their superimposition, in the case of negative chirping,
is easily obtained in such a narrow spectral region. However, an
increase in either the length of the crystal or the chirping
parameter leads to a broadening of the spectrum.

\begin{figure}
\includegraphics[width=0.5\textwidth]{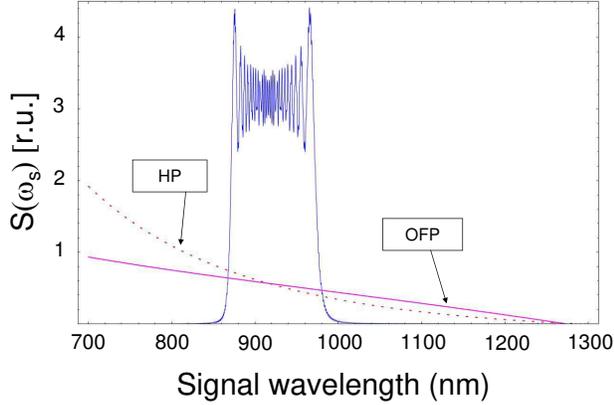}
\caption{(color online) The spectrum of biphotons as well as HP and
OFP frequency dependencies (dashed and solid lines, respectively),
for a crystal with $L = 1.8$ cm and $\alpha = - 50$
cm$^{-2}$}.\label{S5_18}
\end{figure}

\begin{figure}
\includegraphics[width=0.5\textwidth]{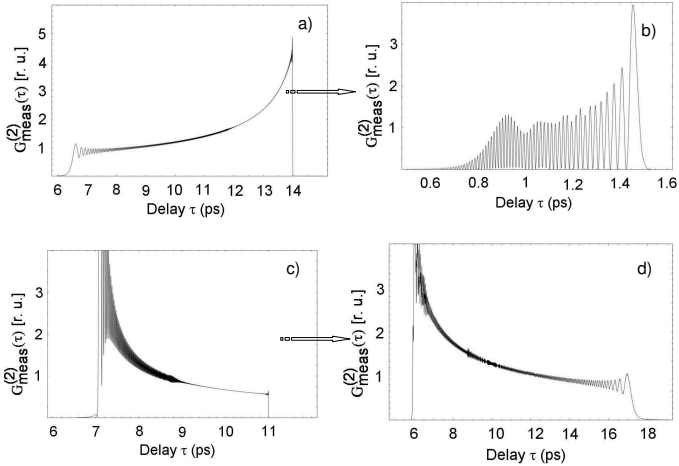}
\caption{(color online) Effect of inserting a 0.52 m long fiber
(right-hand side) on the correlation time for the case shown in Fig.
(\ref{S5_18}) \label{T5_18}}
\end{figure}

\begin{figure}
\includegraphics[width=0.5\textwidth]{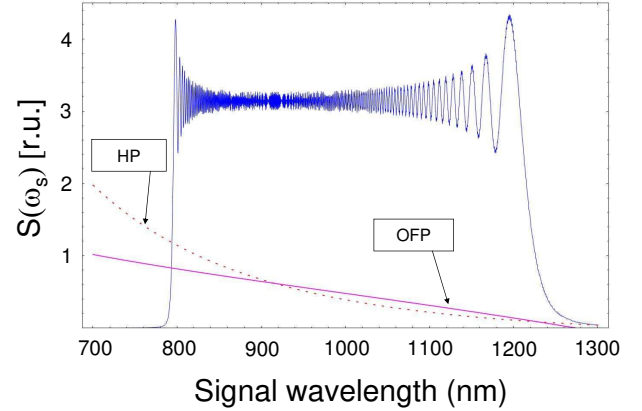}
\caption{(color online) The spectrum of biphotons as well as HP and
OFP frequency dependencies (dashed and solid lines, respectively),
for a crystal with $L = 2.5$ cm and $\alpha = - 50$
cm$^{-2}$}.\label{S5_25}
\end{figure}

Figs. \ref{S5_18} and \ref{S5_25} are analogues of Fig.
\ref{spect_sinc} for two different crystals. In both cases, the same
chirping parameter $\alpha = - 50$ cm$^{-2}$ is considered while the
length is $L = 1.8$ cm in the first case and $L = 2.5$ cm in the
second one. As the crystal length is increased, the spectrum becomes
broader and broader, its width growing from 300 nm in the first case
to 500 nm in the second one. As a consequence, the HP and OFP
frequency dependencies must be superimposed in a larger region, and,
as in general this cannot be achieved, the perfect temporal
compression of the $G^{(2)}_{meas}(\tau)$ is slightly deteriorated.

\begin{figure}
\includegraphics[width=0.5\textwidth]{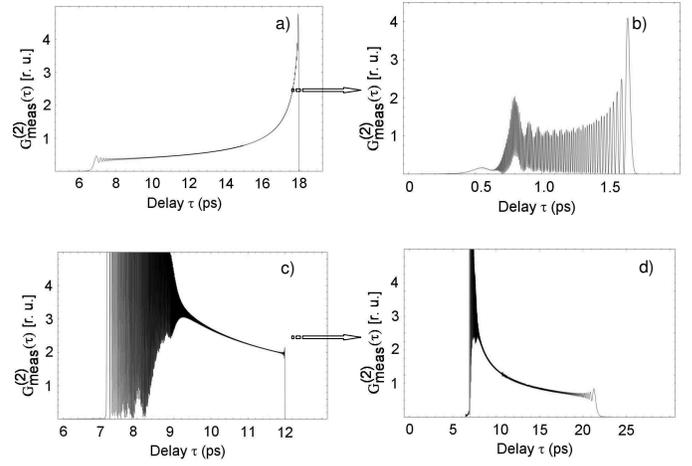}
\caption{(color online) Effect of inserting a 0.57 m long fiber
(right-hand side) on the correlation time for the case shown in Fig.
\ref{S5_25}\label{T5_25}}
\end{figure}

The compressed peaks become more noisy and start to differ from the
sinc function and, as a consequence, the width is slightly
increased. Figs. \ref{T5_18} and \ref{T5_25} show the effect of the
insertion of the fiber. The length of the fiber is 0.52 m in the
first case and 0.57 m in the second case. However, a significant
compression still remains: from 7.5 ps to 700 fs in the first case
and from 11.25 ps to 1 ps in the second case. The correlation time
is reduced in both cases by about an order of magnitude, confirming
that ultrabroadband PDC light can be generated in chirped crystals
and that temporal compression can be achieved in a standard optical
fiber.

\begin{figure}
\includegraphics[width=0.5\textwidth]{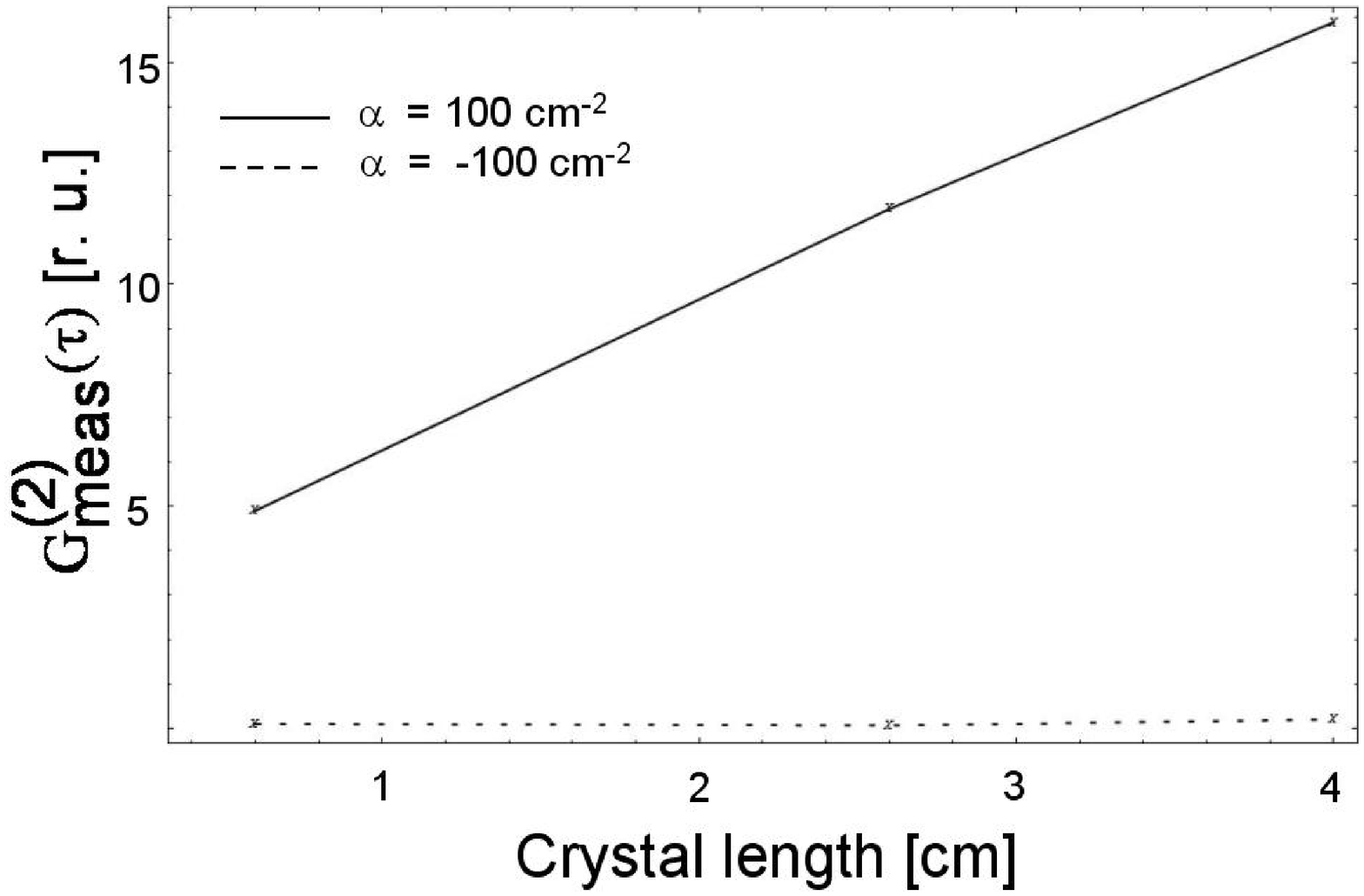}
\caption{(color online) Correlation time as a function of the length
L of the crystal with the chirping parameter $\alpha=100 cm^{-2}$
(solid line) and $\alpha=-100 cm^{-2}$ (dashed line). \label{plot1}}
\end{figure}

Fig. \ref{plot1} shows the correlation time as a function of the
length L of a crystal with the chirping parameter $\alpha=100
cm^{-2}$ (solid line) and $\alpha=-100 cm^{-2}$ (dashed line). Each
point refers to the value that the correlation time takes after the
passage of the signal photon through a fiber with the length
optimized for the maximum compression in the case of
negative-chirped crystals. The plot demonstrates that the difference
between the correlation times for positive and negative chirping
parameters becomes more pronounced for longer crystals. In fact, an
increase in the crystal length leads to a growth of the correlation
time in the case of $\alpha > 0$. Calculation shows that for $\alpha
<0 $, an increase in $L$ affects the correlation time very little,
although a perfect compression would result in the correlation time
inversely proportional to $L$. In any case, the width of compressed
$G_{meas}^{(2)}(\tau)$ is always below 200 fs, at least for the
range of crystal length considered.

\begin{figure}
\includegraphics[width=0.5\textwidth]{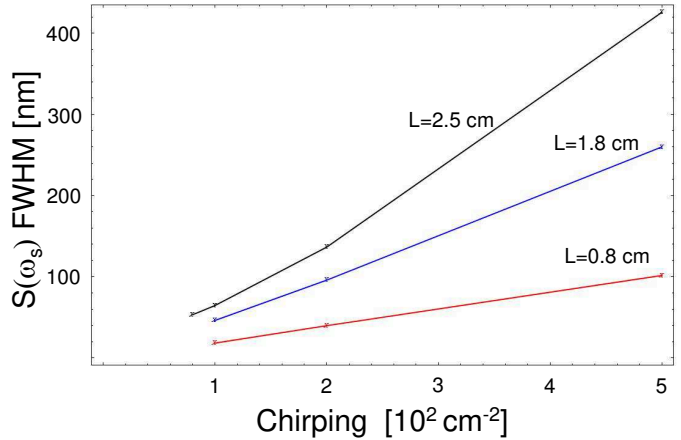}
\caption{(color online) Correlation time as a function of the
chirping parameter, for three different crystal lengths.
\label{plot2}}
\end{figure}

It is interesting to point out that the curves in Fig. \ref{plot1}
are not affected  by changing the absolute value of the chirping
parameter. We have computed the correlation times for both positive
and negative $\alpha$ with the absolute value varying in the range
20 -- 500 $cm^{-2}$, for three lengths of the crystal (0.8 cm, 1.8
cm and 2.5 cm) and no significant difference was observed. This
seems very promising for the temporal compression of ultra-broadband
biphotons, as increasing the chirping parameter leads to a
broadening of the spectrum, as shown in Fig. \ref{plot2}. For
example, for a crystal of $L=2.5$ cm, if the absolute value of the
chirping parameter is increased from 50 cm$^{-2}$ to 500 cm$^{-2}$
the corresponding spectrum width will change from 53.28 nm to 425.86
nm and, at the same time, the correlation time of the compressed
biphoton wavepacket can be still below 1 ps simply with a standard
fiber in the signal channel.

Furthermore, we found that the length of the fiber to be inserted
for the temporal compression is almost independent of the crystal
length. At the same time, it is reduced with the increase of
$\alpha$, passing from 2 m to 50 cm approximately in the range of
the chirping parameters considered above.

\section{\label{sec:level5}Spectral filtering effects}

\begin{figure}
\includegraphics[width=0.5\textwidth]{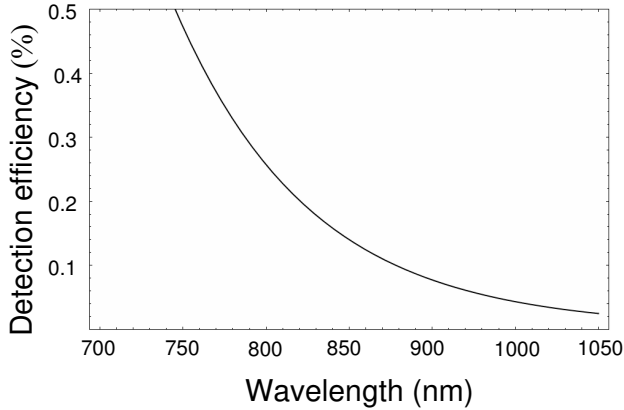}
\caption{(color online) Simulated spectral response of a
single-photon detector. \label{det}}
\end{figure}

In the previous section we considered the spectral and temporal
properties assuming the absence of spectral selection in the
detection arms. Here, we discuss how the presence of such spectral
selection associated with the detection process modifies the
measured spectral and temporal properties of the biphoton wavepacket
as well as the measurement of its temporal compression. In
particular, we consider detection arms with Gaussian spectral
filtering as the one in Eq. (\ref{gauss_fil}) centered at $532$ nm
($\omega_{sc}=\omega_{ic}= 3222 $ THz) and width $250$ nm ($\delta
\omega= 716$ THz). Fig. \ref{det} presents the effective spectral
behavior of such detection arms for signal/idler wavelengths in the
region of interest, and it is in agreement with the typical spectral
response of certain commercial single-photon detectors.

\begin{figure}
\includegraphics[width=0.5\textwidth]{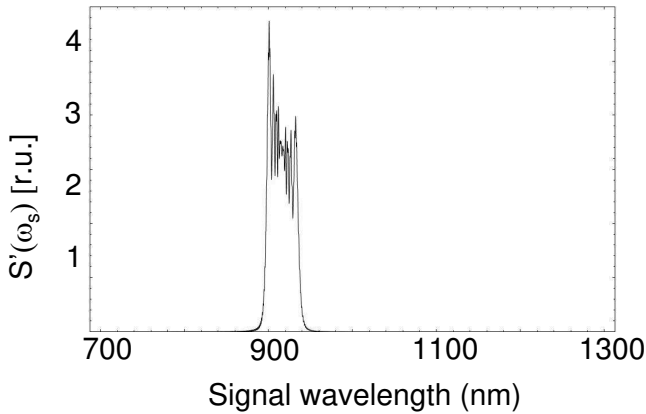}
\caption{(color online) The modified spectrum of biphotons for a
crystal with $L = 0.8$ cm and $\alpha = - 20$
cm$^{-2}$.\label{S2_8_det}}
\end{figure}

\begin{figure}
\includegraphics[width=0.5\textwidth]{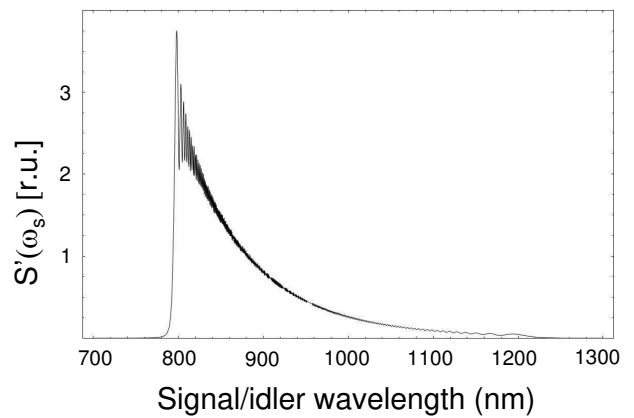}
\caption{(color online) The modified spectrum of biphotons for a
crystal with $L = 2.5$ cm and $\alpha = - 50$
cm$^{-2}$.\label{S5_25_det}}
\end{figure}

For comparison, we analyze the case of two APKTP crystals already
considered in the previous section. The first one has the chirping
parameter $\alpha= -20$ cm$^{-2}$ and length $L=0.8$ cm. The
spectrum not accounting for detection spectral filtering is shown in
Fig. \ref{spect_sinc}, and the corresponding temporal behavior is
shown in Fig. \ref{sinc}. The second one has parameters $\alpha=
-50$ cm$^{-2}$ and $L=2.5$ cm, and its spectral and temporal
behavior (with no account for the detection spectral filtering) is
shown in Fig. \ref{S5_25} and Fig. \ref{T5_25}, respectively. Figs.
\ref{S2_8_det} and \ref{S5_25_det} show the spectrum of the biphoton
wavepackets modified by the detection spectral filtering of Fig.
\ref{det} in the case of the first and the second crystals,
respectively.

\begin{figure}
\includegraphics[width=0.5\textwidth]{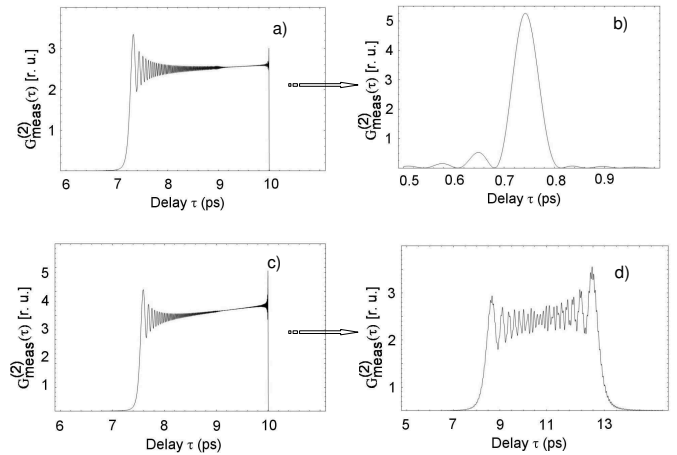}
\caption{(color online) Effect of a 1.06 m long fiber inserted into
the signal channel (right hand side) on $G_{meas}^{(2)}(\tau)$  for
the case shown in Fig. \ref{S2_8_det} \label{T2_8_det}}
\end{figure}

As the spectrum in Fig. \ref{S2_8_det} is quite narrow and its shape
is not modified much with respect to the one in Fig.
\ref{spect_sinc}, it is reasonable to expect that also the
corresponding $G_{meas}^{(2)} (\tau)$, both with and without the
fiber for temporal compression, is not modified much by the presence
of the detection spectral filtering. This is confirmed by the
comparison between Fig. \ref{T2_8_det} and Fig. \ref{sinc}.

\begin{figure}
\includegraphics[width=0.5\textwidth]{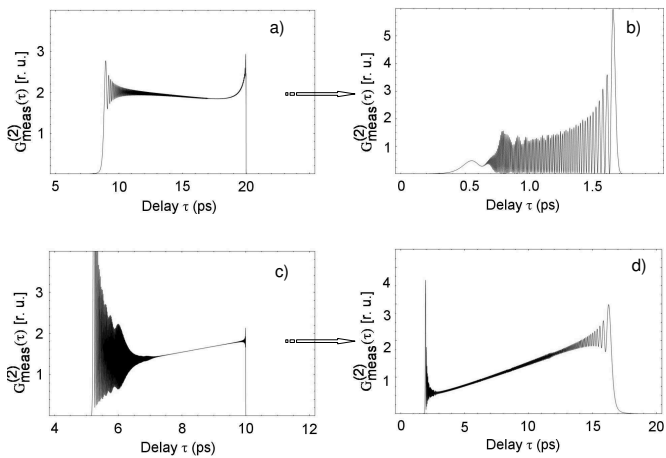}
\caption{(color online) Effect of inserting a 0.57 m long fiber
(right-hand side) on $G_{meas}^{(2)}(\tau)$  for the case shown in
Fig. \ref{S5_25_det}\label{T5_25_det}}
\end{figure}

On the contrary, the wide spectrum of Fig. \ref{S5_25} is strongly
distorted by the presence of the detection spectral filtering as
shown in Fig. \ref{S5_25_det}. The temporal counterpart of this
distortion can be found in Fig. \ref{T5_25_det} where
$G_{meas}^{(2)} (\tau)$ manifests shapes modified with respect to
the ones in Fig. \ref{T5_25}, even if the correlation time widths
look not so different.

Summarizing, the spectral filtering in the detection arms affects
not only the measured spectral properties of the biphoton
wavepacket, but also its measured temporal properties. Thus, to
achieve temporal compression with ultrabroadband PDC light, one has
to take care of the spectral filtering of the detection systems for
its effect on both spectral and temporal behavior of biphoton
wavepackets.

\section{\label{sec:concl}Conclusion}

Producing nonclassical light with tailored spectral and temporal
properties represents a fundamental resource for developing quantum
technologies, such as quantum lithography, quantum metrology,
two-photon coherent absorption, etc. Aiming at an effective
practical realization of our ideas, first presented in
\cite{Nostro}, in this paper we have systematically discussed the
possibility of generating two-photon wavepackets with correlation
times as small as possible by exploiting aperiodically poled
crystals together with standard optical fibers. The analysis carried
out here is numerical, taking into account the exact dispersion
dependencies in the crystal and the fiber, given by the Sellmeier
equations. This is a considerable step forward compared to the
previous work where the effect was mainly described analytically,
but only in the framework of the second-order approximation in the
frequency expansion of the phase mismatch. From our numerical
calculations it follows that higher-order terms in the frequency
expansion of the wavevector mismatch, as well as in the fibre
dispersion dependence, considerably change the shape of the
compressed two-photon wavepacket. However, the effect of compression
is still present. The dependence of this effect on the crystal
length and the chirping parameter has been analyzed. Our results
clearly show that a significant compression, i.e. reduction of
correlation times by more than one order of magnitude, can be easily
reached, achieving correlation times as short as hundreds of
femtoseconds. In our analysis we also discussed the effect of
possible spectral filtering in detection arms. Despite spectral
filtering modifies the correlation time of the biphoton wavepacket
as well as its spectrum, it is noteworthy to observe that, at least
in the case considered, the temporal compression is essentially
preserved. Direct experimental observation of such a compression by
means of coincidence measurement is possible only if the detectors'
jitter is smaller than the correlation time of the biphoton
wavepacket. Even if hundreds of femtoseconds jitter seems hardly
achievable for present days detector technologies, it could be a
target of next future detectors \cite{det1, det2, det3, det4}.
Furthermore, undirect measurements can be envisaged \cite{wip}.
Since our scheme only requires relatively simple and available
technologies (i.e. chirped poled crystals and commercial optical
fibers), we have demonstrated the feasibility of a suitable
 practical realization of ultracompressed biphotons.

\begin{acknowledgments}
This work has been supported in part by MIUR (PRIN 2007FYETBY), "San
Paolo foundation", NATO (CBP.NR.NRCL 983251), RFBR 08-02-00555a,
Russian Federal Agency for Science and Innovation (Rosnauka) state
contract 02.740.11.0223, and the Russian Program for Scientific
Schools Support, grant \# NSh-796.2008.2. M.~V.~Ch. also
acknowledges the support of the CRT Foundation.
\end{acknowledgments}

\end{document}